\documentclass[%
pre,preprint,superscriptaddress,
tightenlines,
showpacs,
showkeys,
a4paper,12pt
]{revtex4}
\usepackage{amssymb,amsmath}
\usepackage{graphicx}

 \newcommand{\T}{\textit{T}}
 \newcommand{\bs}[1]{\boldsymbol{#1}}
 \newcommand{\vc}[1]{\mathbf{#1}}
 \newcommand{\uvc}[1]{\mathbf{\hat #1}}

\begin{document}
 \DeclareGraphicsExtensions{.jpg,.pdf}

 \title{\bfseries Light scattering by optically anisotropic scatterers
II: \\ \T--matrix computations for radially and uniformly anisotropic
droplets}

\author{A.D.~Kiselev}
\email[E-mail: ]{kisel@elit.chernigov.ua}
\affiliation{%
 Chernigov State Technological University,
 Shevchenko Street 95,
 14027 Chernigov, Ukraine
} 
\author{V.Yu.~Reshetnyak}
\email[E-mail: ]{reshet@iop.kiev.ua}
\affiliation{%
 Kiev University,
 Prospect Glushkova 6,     
 03680 Kiev, Ukraine
} 
\author{T.J.~Sluckin}
\email[E-mail: ]{t.j.sluckin@maths.soton.ac.uk} 
\affiliation{%
 Faculty of Mathematical Studies, 
 University of Southampton,
 Southampton, SO17 1BJ, UK
 }

\date{\today}

 \begin{abstract}
This is the second paper in a series on light scattering from
optically anisotropic scatterers embedded in an isotropic medium. The
apparently complex T-matrix theory involving mixing of angular
momentum components turns out to be an efficient approach to
calculating scattering in these systems.  We present preliminary
results of numerical calculations of the scattering by spherical
droplets in some simple cases. The droplets contain optically
anisotropic material with local radial or uniform anisotropy.  We
concentrate on cases in which the scattering is due only to the local
optical anisotropy within the scatterer.  For radial anisotropy we
find non-monotonic dependence of the scattering cross-section on the
degree of anisotropy can occur in a regime for which both the Rayleigh
and semi-classical theories are inapplicable.  For uniform anisotropy
the cross-section is strongly dependent on the angle between the
incident light and the optical axis, and for larger droplets this
dependence is non-monotonic.
\end{abstract}

\pacs{%
42.25.Fx, 77.84.Nh
}

\keywords{%
light scattering; anisotropy; \T-matrix theory
}

\maketitle

\section{Introduction}
\label{sec:intro}
In the first paper in this series~\cite{Kis:2001}, we developed a
\T-matrix theory of light scattering by spherical but optically
anisotropic -- either radially or uniformly -- scatterers. Whereas for
radial anisotropic scatterers it is possible to derive closed form
expressions for the elements of the T-matrix, in the uniformly
anisotropic case this is no longer true. To further complicate
matters, in this latter case the \T--matrix elements can only be
derived numerically by inverting a rather difficult set of matrix
equations. Because the spherical symmetry of a usual scattering
problem is reduced in this case to cylindrical symmetry, the
scattering involves angular momentum mixing.  The consequence is that
the set of equations to be inverted is in principle infinite in
number.

In this paper we continue this study. We find that notwithstanding the
daunting nature of the formal structure of the \T--matrix theory, in
fact this theory provides a viable and efficient strategy for
calculating the elements of the \T--matrix, and thus computing the
scattering properties of these objects.  Throughout this paper we
shall refer to the first paper in this series as I, and write Eq.~(n)
of I as Eq.~(n.I).

A number of approaches are available to study light scattering by
complex objects.  A brief summary is as follows. The scattering
amplitudes can be described using Green's function
techniques~\cite{New}, but these involve solving complex integral
equations over infinite domains.  Under some circumstances one can
approximate the kernels of these equations either as the incident wave
or as a semi-classical perturbed wave, leading to the well-known
Rayleigh-Gans (R-G) and Anomalous Diffraction Approximations (ADA).
These have been used by \v{Z}umer and coworkers to examine the
problems we consider in this paper~\cite{Zum:1986,Zum:1988}. However,
the approximations are only valid over certain wavelength and optical
contrast regimes.  The century-old Mie strategy and its modern
\T--matrix extensions yield exact solutions, but unfortunately this
approach does not work in every case. Finally one can of course use
real space finite element approaches, but these are notoriously
inefficient at reproducing known solutions. For a more comprehensive
review we refer the reader to Chap.~2 in~\cite{Mish} and references
therein.
 
The \T-matrix theory \emph{is} known to be a computationally efficient
approach to study light scattering by nonspherical optically isotropic
particles~\cite{Mish}.  One may thus expect that a \T--matrix approach
to geometrically spherical but optically non-spherical scatterers can
at the very least enable scattering properties to be evaluated when
the approximate methods cannot be applied. In addition, whereas the
region of validity of the approximate methods such as R-G and ADA in
the case of isotropic scatterers is reasonably well-understood, in the
case of anisotropic scatterers this problem has not been been studied
in any detail.

In I we have discussed composite spherical scatterers, consisting of a
central isotropic core plus a surrounding annular layer in which the
optical tensor is anisotropic: 
$\bs{\epsilon}_{ij} =\epsilon_{\perp}[
\delta_{ij}+ u (\uvc{n}\otimes\uvc{n})_{ij}] $, where 
$u = (\epsilon_{\parallel}-\epsilon_{\perp})/\epsilon_{\perp}$ is the
anisotropy parameter. For radial anisotropy the optical axis,
$\uvc{n}$, is directed along the radius vector, $\uvc{n}=\uvc{r}$, and
for the uniform anisotropy the optical axis is parallel to the
$z$-axis, $\uvc{n}=\uvc{z}$. These cases present different
mathematical challenges to the theorist.

Light scattering from the radially anisotropic annular layer was first
studied long ago by Roth and Digman~\cite{Rot:1973} using the
technique normally known as Debye potentials.  In an earlier
paper~\cite{Kis:2000:opt} we have recovered this solution as a special
case of a more general set of anisotropies. A crucial step in the
derivation of this result involves writing the so-called modified
\T-matrix ansatz (see Eqs.~(26.I)-(27.I)).  The spherical symmetry of
the scatterer requires the modes~(27.I) entering the ansatz~(26.I) to
be proportional to the corresponding vector spherical harmonics. As a
result the scattering does not mix different angular momenta. The
\T--matrix is then diagonal over the angular momentum indices and the
azimuthal numbers and the elements of the \T-matrix are expressible in
closed form (see Sec.~IV.A of~I).

The uniform anisotropy case is much more difficult. The light
scattering problem for a uniformly anisotropic spherical scatterer is
not exactly solvable~\cite{Bor}. As an alternative, it has been studied by
using R-G and ADA ~\cite{Zum:1986,Zum:1988}.

In I we approached this problem by formulating it as a suitably
modified \T-matrix theory. It is necessary to relate the plane wave
packet representation to expansions of electromagnetic fields over
vector spherical harmonics. The net result is to define a set of wave
functions representing \emph{exact solutions} of Maxwell's equations
in an anisotropic layer that are at the same time deformed spherical
harmonics.  The coefficient functions that enter the expressions for
these wave functions describe angular momentum mixing and are
computationally accessible.

In this paper we find numerical results for the total scattering
cross-section in the limiting case of a droplet, i.e. when the radius
of the isotropic core of the scatterer, $R_2$, is negligible 
($R_2\to 0$). The scattering by a droplet presents fewer technical 
difficulties than scattering by the annular layer which was our
principal focus in I.  Anisotropy effects are our primary concern and
for this reason we pay special attention to the case where the
scattering can be solely attributed to the presence of the anisotropy.
Specifically, in our subsequent calculations we consider the special
case for which the ordinary wave refractive index and the refractive
index of the material are matched, 
$n=n_o$ or $\epsilon=\epsilon_{\perp}$.

The paper is organised as follows. In Sec.~\ref{sec:t-matrix} we use
the general \T-matrix formalism developed in I to write down some
necessary formulae relevant to the special case of droplets. In
Sec.~\ref{sec:num-res} we make brief comments on the numerical
strategy and present some numerical results.
Finally in Sec.~\ref{sec:concl} we make some brief concluding remarks.

\section{\T-matrix calculations for droplets}
\label{sec:t-matrix}
\subsection{Notation}

In this section we adapt the key theoretical relations derived in~I 
so that they can be used in the case of droplets.
In addition, we rewrite the expressions for the total scattering
cross-section in a more convenient form.

\begin{figure*}[!thb]
\centering
\resizebox{100mm}{!}{\includegraphics{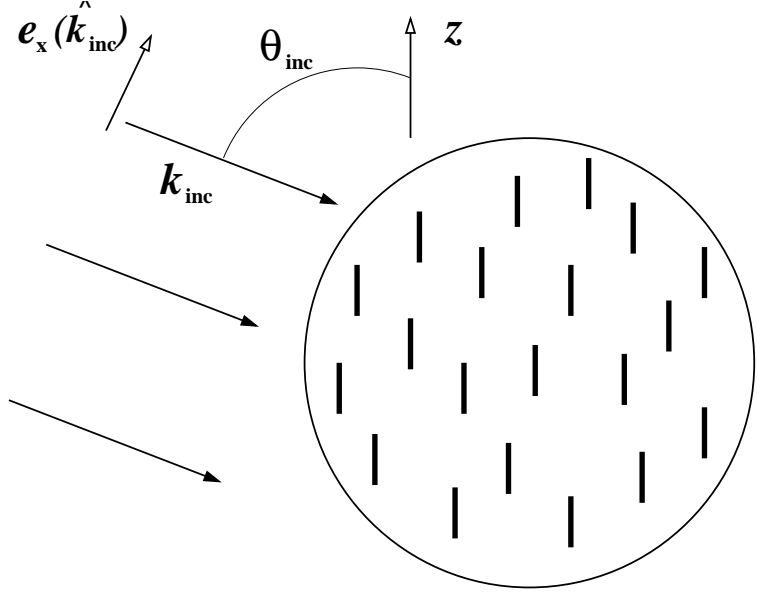}}
\caption{
Uniformly anisotropic droplet.
The polarisation vector $\vc{e}_{y}(\uvc{k}_{inc})$
is normal to the plane of the picture,
$\vc{e}_{y}(\uvc{k}_{inc})\propto\uvc{k}_{inc}\times\uvc{z}$.
Inside the droplet plane waves linearly polarised along
$\vc{e}_{y}(\uvc{k}_{inc})$ represent ordinary waves.}
\label{fig:droplet}
\end{figure*}

We first introduce some notation.  Radially anisotropic droplets
present an isotropic face to the world, to the extent that the
scattering properties are independent of the the direction of the
incident wave.  In the case of the uniformly anisotropic droplets,
this is no longer true.  In this case the scattering geometry is shown
in Fig.~\ref{fig:droplet}.
The angle of incidence $\theta_{\rm inc}$ is the angle between the
direction of incidence $\hat{k}$ and the direction of the uniform
anisotropy.  The $y$ direction is perpendicular to the plane made from
these two directions.  We shall show explicitly that the scattering
process does not involve the $y$-component, $E_{y}^{(inc)}$, of the
incoming plane wave
\begin{widetext}
\begin{equation}
  \label{eq:inc}
\vc{E}_{inc}=\vc{E}^{(inc)}\exp(i\,\vc{k}_{inc}\cdot\vc{r})\, ,
\quad
\vc{E}^{(inc)}=
E_{x}^{(inc)}\vc{e}_{x}(\uvc{k}_{inc})+
E_{y}^{(inc)}\vc{e}_{y}(\uvc{k}_{inc})\, ,
\qquad
\vc{k}_{inc}=k\uvc{k}_{inc}\, ,
\end{equation}
\end{widetext}  
provided the refractive indices $n\equiv\sqrt{\epsilon}$ and 
$n_o\equiv\sqrt{\epsilon_{\perp}}$ are matched.
In this paper we put the magnetic permittivity
equal to the unit and slightly change the notations: 
$n_0$ and $\epsilon_{\perp}$ instead of $n_1$ and $\epsilon_1$.  
This corresponds to the physical condition in which there is 
a matching condition for the refracted 
ordinary wave inside the scattering droplet.

\subsection{Equations for \T-matrix}
\label{subsec:eq-t-matr}


Since the electromagnetic field must be regular at the origin, the
harmonics inside the droplet are now given by
\begin{widetext}
\begin{subequations}
\label{eq:core}
\begin{align}
& \vc{E}_{jm} =
\alpha_{jm}^{(c)}\,\vc{P}^{(m)}_{jm}(\rho_o,\uvc{r})
-n_o^{-1}\,
\tilde\alpha_{jm}^{(c)}\,\vc{P}^{(e)}_{jm}(\rho_o,\uvc{r})\, ,
\label{eq:core_e}\\
& \vc{H}_{jm} =
\tilde\alpha_{jm}^{(c)}\,\vc{Q}^{(m)}_{jm}(\rho_o,\uvc{r})+
n_o\,
\alpha_{jm}^{(c)}\,\vc{Q}^{(e)}_{jm}(\rho_o,\uvc{r})\, ,
\label{eq:core_h}
\end{align} 
\end{subequations}
where $\rho_o= m_o kr=m_o\rho$ and 
$m_o\equiv n_o/n$ is the optical contrast.
\end{widetext}
The modes $\vc{P}^{(\alpha)}_{jm}$
and $\vc{Q}^{(\alpha)}_{jm}$  
have been given by Eqs.~(27.I) for radial anisotropy and 
by Eqs.~(43.I) for uniform anisotropy.

Following Eqs.~(32.I), the continuity conditions at the outside of the
droplet, $r=R_1$, can then be written in matrix notation as follows:
\begin{widetext}
\begin{equation}
\sum_{j'\ge |m|}\vc{R}_1^{jj';\,m}
\begin{pmatrix}
\alpha_{j'm}^{(c)}\\
\tilde{\alpha}_{j'm}^{(c)}
\end{pmatrix}
 =
\bs{\Gamma}_1^{\,j}
\begin{pmatrix}
\alpha_{jm}^{(inc)}\\
\tilde{\alpha}_{jm}^{(inc)}
\end{pmatrix}
+
\tilde{\bs{\Gamma}}_1^{\,j}
\begin{pmatrix}
\beta_{jm}^{(sca)}\\
\tilde{\beta}_{jm}^{(sca)}
\end{pmatrix}\, ,
\label{eq:sys}
\end{equation}

\begin{equation}
\bs{\Gamma}^{\,j}(r)=
\begin{pmatrix}
j_j(\rho) & 0\\
n [j_j(\rho)]' & 0\\
 0 & j_j(\rho)\\
0 & -n^{-1} [j_j(\rho)]'
\end{pmatrix}\, ,\quad
\tilde{\bs{\Gamma}}^{\,j}(r)=
\begin{pmatrix}
h_j^{(1)}(\rho) & 0\\
n [h_j^{(1)}(\rho)]' & 0\\
 0 & h_j^{(1)}(\rho)\\
0 & -n^{-1} [h_j^{(1)}(\rho)]'
\end{pmatrix}\, ,
\label{eq:gamma}
\end{equation}
\end{widetext}
where the index 1 indicates that matrix elements are calculated
at the boundary of droplet, $r=R_1$.

In the case of a radially anisotropic droplet the matrix on the left
hand side of Eq.~\eqref{eq:sys} is diagonal over angular momentum
numbers
\begin{widetext}
\begin{equation}
\vc{R}^{jj';\,m}(r) = \delta_{jj'}
\begin{pmatrix}
j_j(\rho_o) & 0\\
n_o [j_j(\rho_o)]' & 0\\
 0 & j_{\tilde{j}}(\rho_o)\\
0 & -n_o^{-1} [j_{\tilde{j}}(\rho_o)]'
\end{pmatrix}\, ,\quad 
\tilde{j}(\tilde{j}+1)=j(j+1)/(1+u)\, .
\label{eq:r_rad}
\end{equation}
\end{widetext}

When the droplet is uniformly anisotropic, the matrix $\vc{R}^{jj';\,m}(r)$ 
is no longer diagonal over angular momentum quantum numbers. In this
case it takes the form:
\begin{widetext}
\begin{equation}
\vc{R}^{jj';\,m}(r)= 
\begin{pmatrix}
p_{jj';\,m}^{\,(m,m)}(\rho_o) 
& -n_o^{-1}\,p_{jj';\,m}^{\,(m,e)}(\rho_o) \\
n_o\,q_{jj';\,m}^{\,(e,e)}(\rho_o) 
& q_{jj';\,m}^{\,(e,m)}(\rho_o)\\
n_o\,q_{jj';\,m}^{\,(m,e)}(\rho_o) 
& q_{jj';\,m}^{\,(m,m)}(\rho_o)\\
p_{jj';\,m}^{\,(e,m)}(\rho_o) 
& -n_o^{-1}\,p_{jj';\,m}^{\,(e,e)}(\rho_o)
\end{pmatrix}\, ,
\label{eq:r_uni}
\end{equation}
\end{widetext}
where the coefficient functions are given 
by Eqs.~(C3.I)-(C10.I) in Appendix~C of~I.

The system~\eqref{eq:sys} can be then simplified by multiplying
both sides by the matrices
\begin{widetext}
\begin{subequations}
\label{eq:aux_matr}
\begin{align}
&
\vc{H}^{\,j}(r)= -i\,\rho^{2}\,
\begin{pmatrix}
[j_j(\rho)]'& -n^{-1}\,j_j(\rho)& 0 & 0\\
0 & 0 &n^{-1}\,[j_j(\rho)]'& j_j(\rho)
\end{pmatrix}\, ,
\\
&
\tilde{\vc{H}}^{\,j}(r)= -i\,\rho^{2}\,
\begin{pmatrix}
[h_j^{(1)}(\rho)]'& -n^{-1}\,h_j^{(1)}(\rho)& 0 & 0\\
0 & 0 &n^{-1}\,[h_j^{(1)}(\rho)]'& h_j^{(1)}(\rho)
\end{pmatrix}\, .
\end{align}
\end{subequations}
\end{widetext}
Using the fact that the
Wronskian for spherical Bessel
functions is given by~\cite{Abr}: 
$$
W\{j_j(\rho),h_j^{(1)}(\rho)\}=i/\rho^2\, ,
$$
we derive a system equivalent to Eqs.~\eqref{eq:sys} in the
following form: 
\begin{widetext}
\begin{subequations}
\label{eq:sys_fin}
\begin{align}
&
\begin{pmatrix}
\alpha_{jm}^{(inc)}\\
n^{-1}\,\tilde{\alpha}_{jm}^{(inc)}
\end{pmatrix}
=
\sum_{j'\ge |m|}\vc{B}_{jj';\,m}
\begin{pmatrix}
\alpha_{j'm}^{(c)}\\
\tilde{\alpha}_{j'm}^{(c)}
\end{pmatrix}\, ,
\label{eq:sys_inc}\\
&
\begin{pmatrix}
\beta_{jm}^{(sca)}\\
n^{-1}\,\tilde{\beta}_{jm}^{(sca)}
\end{pmatrix}
=
\sum_{j'\ge |m|}\vc{A}_{jj';\,m}
\begin{pmatrix}
\alpha_{j'm}^{(c)}\\
\tilde{\alpha}_{j'm}^{(c)}
\end{pmatrix}\, ,
\label{eq:sys_sca}
\end{align}
\end{subequations}
where $\vc{A}_{jj';\,m} = - \vc{H}_1^{\,j}\cdot\vc{R}_1^{jj';\,m}$
and $\vc{B}_{jj';\,m} =  \tilde{\vc{H}}_1^{\,j}\cdot\vc{R}_1^{jj';\,m}$.
\end{widetext}

These are the equivalent of the equations (51.I) for scattering by a
spherical annulus.  The equations are considerably simpler, as a
result of what might be called mode decoupling.  The crucial point is
that the normal modes inside the droplet are all regular at the
origin, and thus two types of modes which appear in the case of the
annulus do not appear here.  This reduces the number of variables in
the problem by one half, even in the case of the uniform anisotropy
which, at least in principle, presented a certain number of problems
in the annular case.

From Eq.~\eqref{eq:sys_fin}  
we can immediately derive  an expression for the \T-matrix:
\begin{equation}
\vc{T}_{jj';\,m} = [\vc{A}\cdot\vc{B}^{-1}]_{jj';\,m}\, .
\label{eq:t_matr}
\end{equation}

For radial anisotropy all the matrices  
on the right hand sides of
Eqs.~\eqref{eq:sys_fin}
are diagonal. So, it is easy to write down the result
for the \T-matrix:
\begin{widetext}
\begin{equation}
\vc{T}_{jj';\,m} = \delta_{jj'}
\begin{pmatrix}
T_j^{11} & 0\\
0 & T_j^{22}
\end{pmatrix}\, ,
\label{eq:t_matr_rad}
\end{equation}
where
\begin{subequations}
  \label{eq:t_matr_el_rad}
\begin{align}
&
T_j^{11}=-\frac{
[j_j(\rho_o)]_1\,[j_j(\rho)]'_1-
m_o\,[j_j(\rho_o)]'_1\,[j_j(\rho)]_1}
{[j_j(\rho_o)]_1\,[h_j^{(1)}(\rho)]'_1-
m_o\,[j_j(\rho_o)]'_1\,[h_j^{(1)}(\rho)]_1 
}\, ,
\label{eq:t11_rad}\\
&
T_j^{22}=-\frac{
m_o\,[j_{\tilde{j}}(\rho_o)]_1\,[j_j(\rho)]'_1-
[j_{\tilde{j}}(\rho_o)]'_1\,[j_j(\rho)]_1}
{m_o\,[j_{\tilde{j}}(\rho_o)]_1\,[h_j^{(1)}(\rho)]'_1-
[j_{\tilde{j}}(\rho_o)]'_1\,[h_j^{(1)}(\rho)]_1 
}
\, .
\label{eq:t22_rad}
\end{align}
\end{subequations}
\end{widetext}
These formulae bear close resemblance to the well known Mie
expressions~\cite{Bor}.

\begin{figure*}[!thb]
\centering
\resizebox{150mm}{!}{\includegraphics{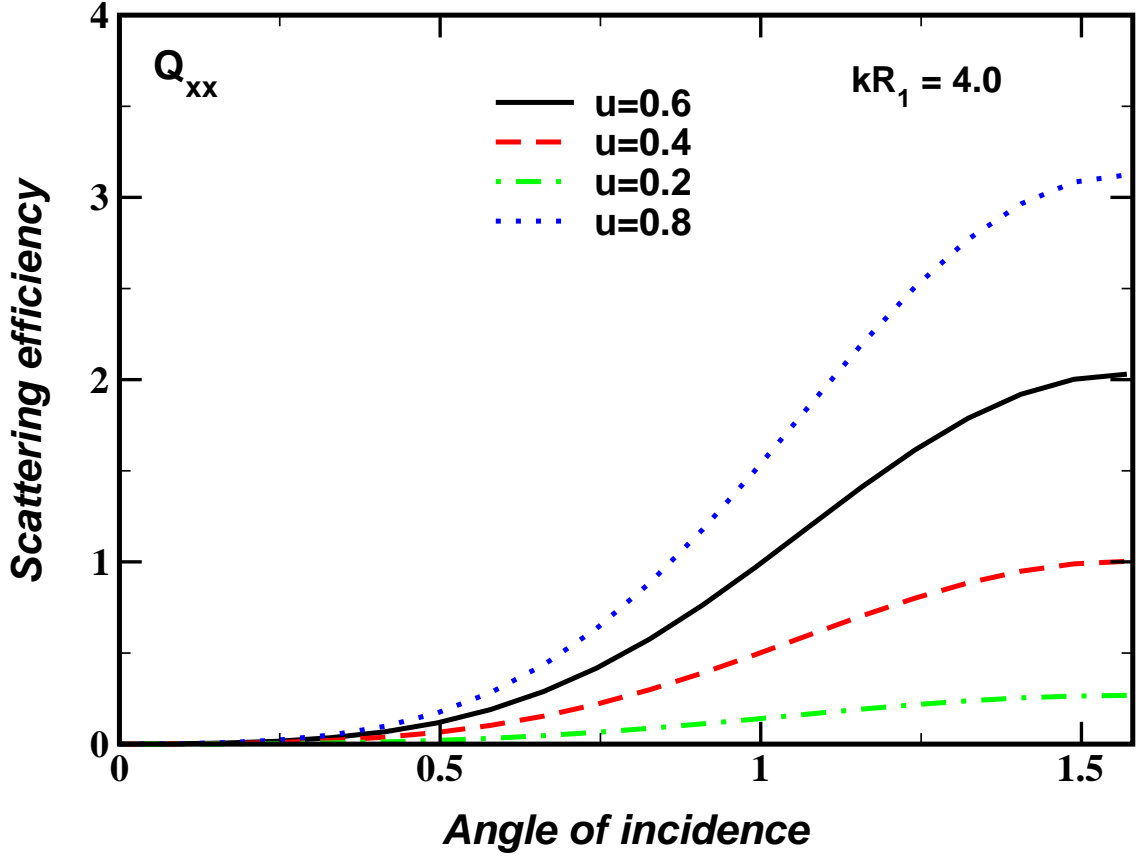}}
\caption{
Scattering efficiency of 
uniformly anisotropic droplet as a function the angle of incidence
(the angle between the incident wave 
and the optical axis)
at various values of the anisotropy parameter, 
$u=(\epsilon_{\parallel}-\epsilon_{\perp})/\epsilon_{\perp}$,
with $kR_1=4.0$
and $n=n_o$.
}
\label{fig:uni_ang_u}
\end{figure*}

\begin{figure*}[!thb]
\centering
\resizebox{150mm}{!}{\includegraphics{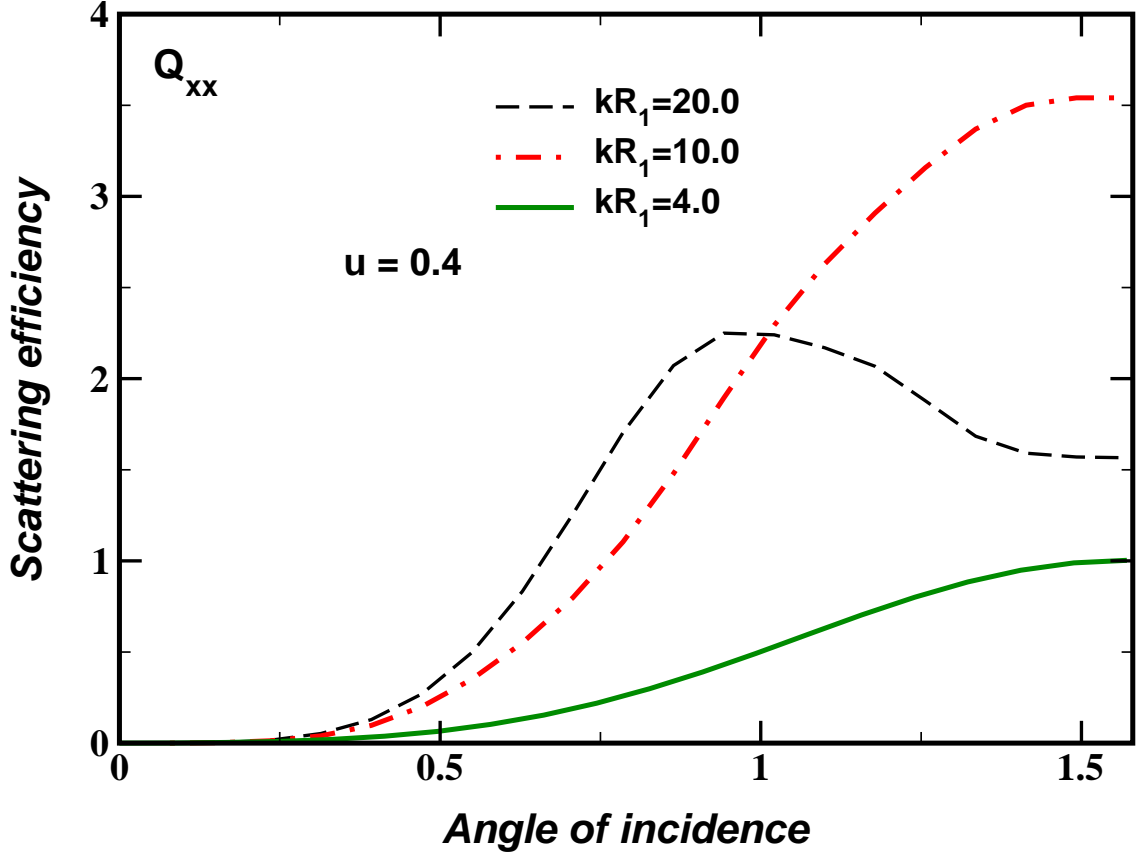}}
\caption{
Dependence of the scattering efficiency 
on the angle of incidence
for uniformly anisotropic droplet 
at various values of the size parameter and $u=0.4$.
The refractive indices $n$ and $n_o$ are matched.
}
\label{fig:uni_ang_kr}
\end{figure*}

\begin{figure*}[!thb]
\centering
\resizebox{150mm}{!}{\includegraphics{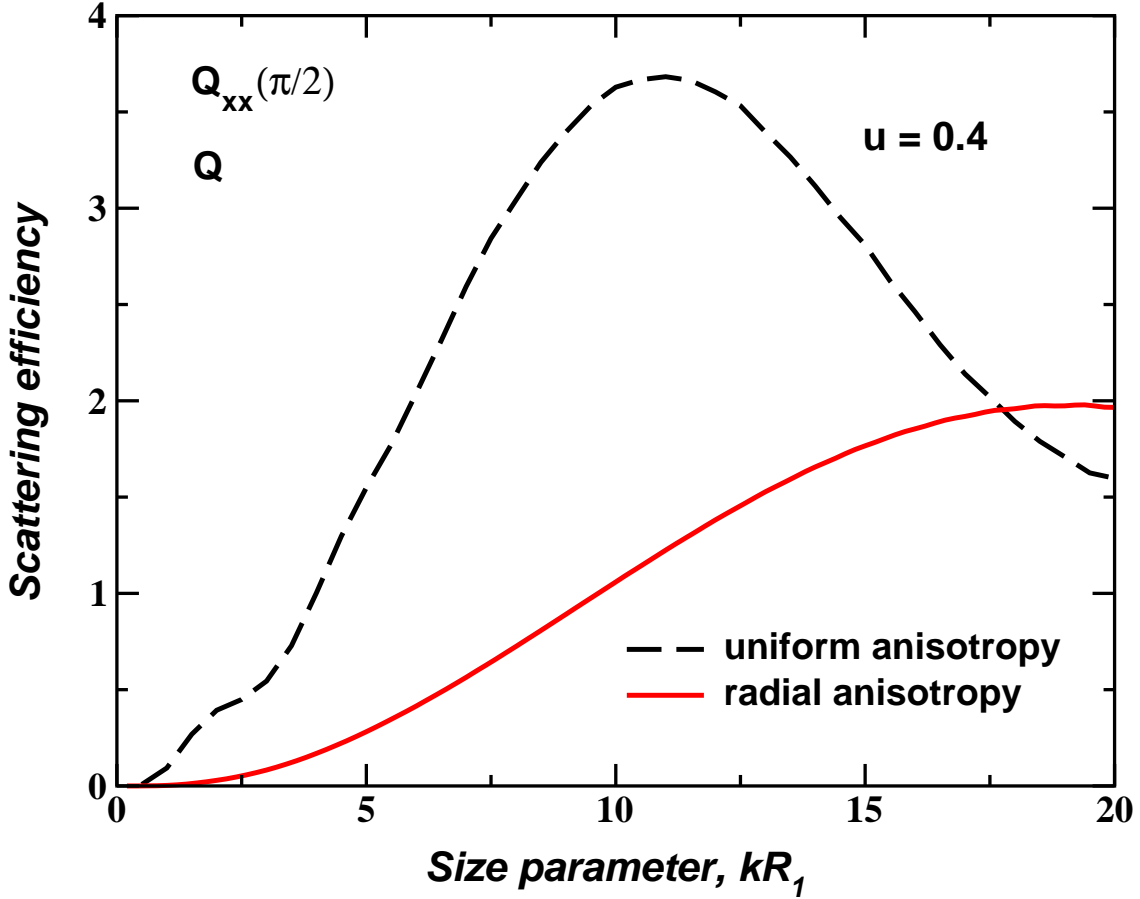}}
\caption{
Scattering efficiencies of 
radially and  uniformly anisotropic droplets 
versus the size parameter 
at $u=0.4$, $\theta_{inc}=\pi/2$ and
$n=n_o$.
}
\label{fig:eff_vs_kr}
\end{figure*}

\begin{figure*}[!thb]
\centering
\resizebox{150mm}{!}{\includegraphics{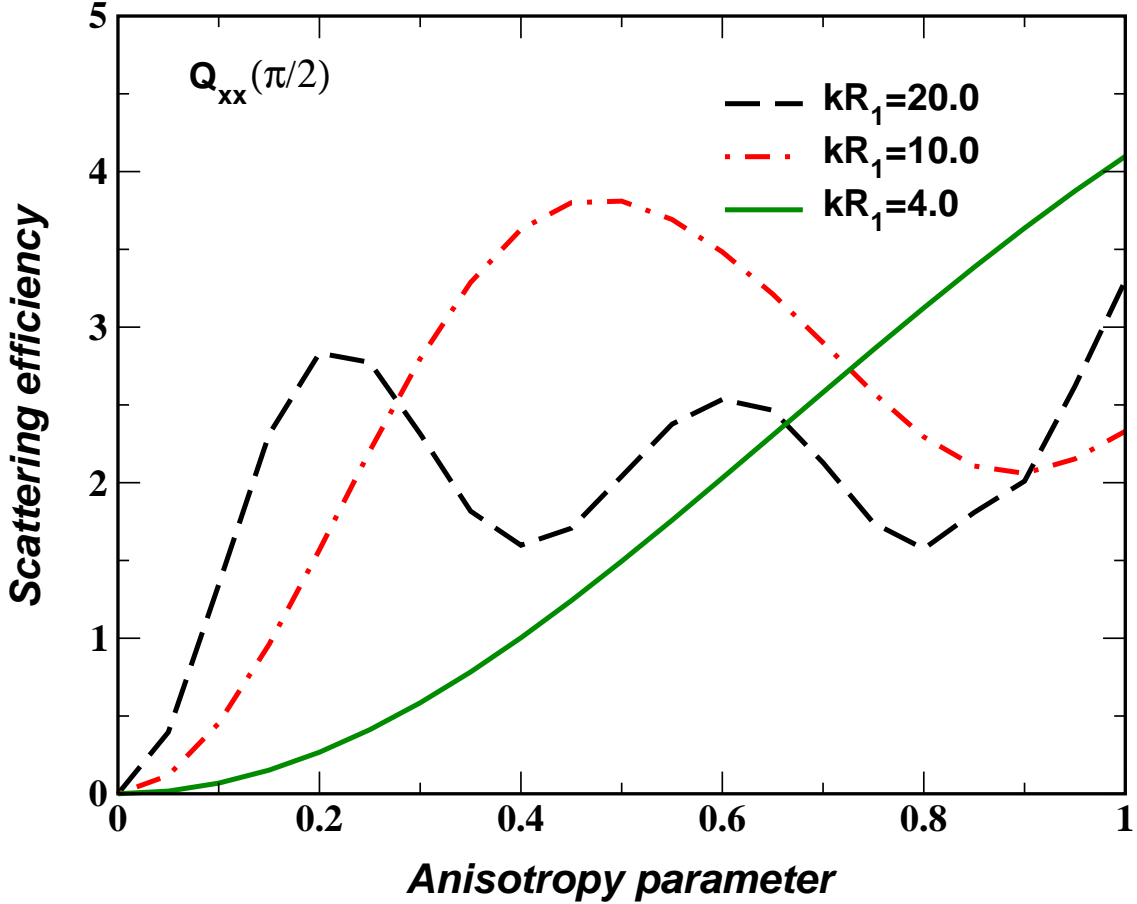}}
\caption{
Scattering efficiencies of 
radially and  uniformly anisotropic droplets 
versus the anisotropy parameter
at various values of the size parameter for 
$\theta_{inc}=\pi/2$
and
$n=n_o$.
}
\label{fig:eff_u_uni}
\end{figure*}

\begin{figure*}[!thb]
\centering
\resizebox{150mm}{!}{\includegraphics{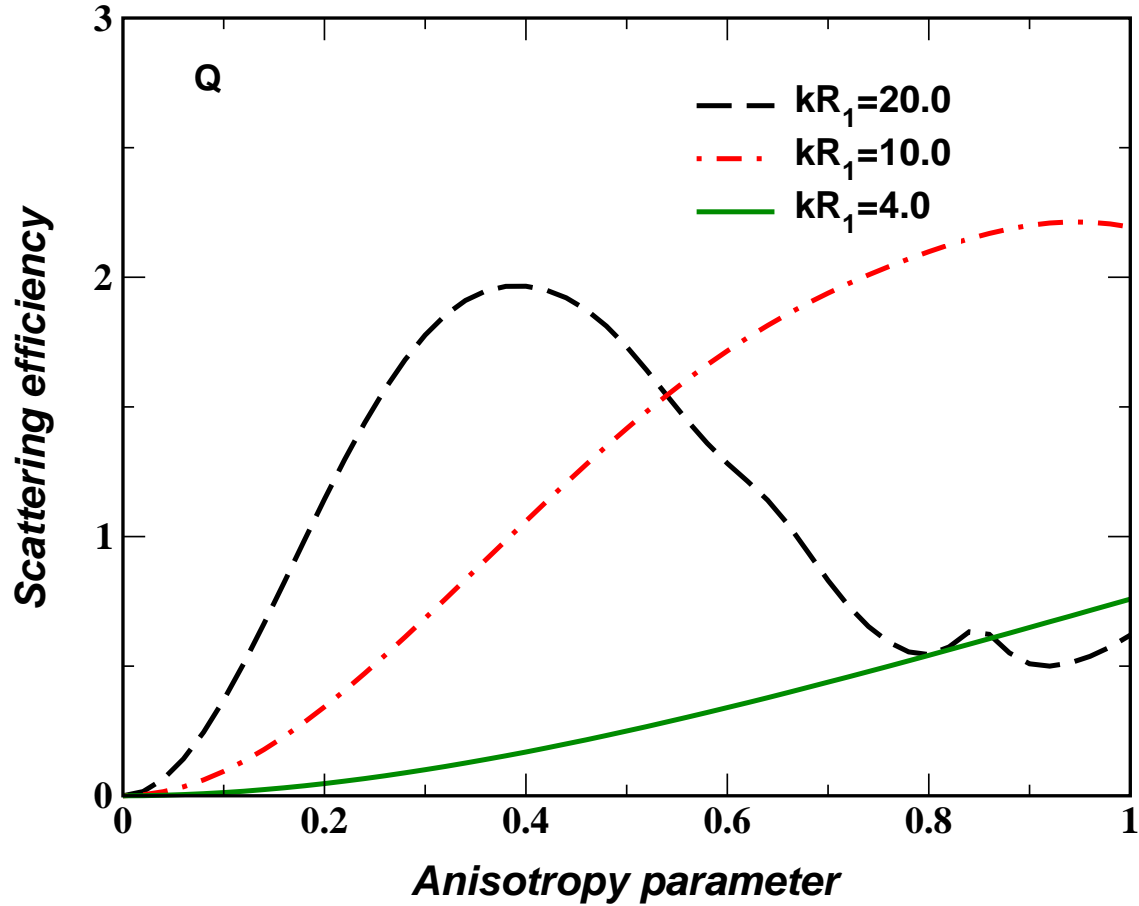}}
\caption{
Dependence of the scattering efficiency 
on the anisotropy parameter
for radially anisotropic droplet 
at various values of the size parameter and $n=n_o$.
}
\label{fig:eff_u_rad}
\end{figure*}

\subsection{Scattering efficiency}
\label{subsec:scatt-eff}

We have seen in~I that the total scattering cross-section for the
uniformly anisotropic scatterer depends on the angle of incidence
$\theta_{inc}$.  The scattering cross-section also depends on the
polarisation of the incident wave.  In order to emphasise this, let us
express the coefficients of the incident wave $\alpha_{jm}^{(inc)}$
and $\tilde{\alpha}_{jm}^{(inc)}$ (see Eq.~(9.I)) as follows
\begin{widetext}
\begin{subequations}
\begin{align}
&
\alpha_{jm}^{(inc)}=\alpha_{jm;\,x}^{(inc)}\,E_x^{(inc)}
+
\alpha_{jm;\,y}^{(inc)}\,E_y^{(inc)}\, ,
\label{eq:alp}\\
&
\tilde{\alpha}_{jm}^{(inc)}=\tilde\alpha_{jm;\,x}^{(inc)}\,E_x^{(inc)}
+
\tilde\alpha_{jm;\,y}^{(inc)}\,E_y^{(inc)}\, ,
\label{eq:talp}
\end{align}
\end{subequations}
where
\begin{align}
& 
\alpha_{jm;\,x}^{(inc)}=i^j\, [\pi(2j+1)]^{1/2}
D_{jm}^{(y)}(\uvc{k}_{inc})\, ,\quad
\alpha_{jm;\,y}^{(inc)}= i^{j+1}\, [\pi(2j+1)]^{1/2}
D_{jm}^{(x)}(\uvc{k}_{inc})\, ,\notag\\
&
\tilde\alpha_{jm;\,x}^{(inc)}=n\, i^{j+1}\, [\pi(2j+1)]^{1/2}
D_{jm}^{(x)}(\uvc{k}_{inc})\, ,\quad
\tilde\alpha_{jm;\,y}^{(inc)}=-n\,i^j\, [\pi(2j+1)]^{1/2}
D_{jm}^{(y)}(\uvc{k}_{inc})\, .
\end{align}
\end{widetext}
From Eq.~(19.I) 
the scattering efficiency then can be rewritten in the following
form:
\begin{widetext}
\begin{gather}
Q=\frac{S_{sca}}{\pi R_1^2}=I_{inc}^{-1}
\sum_{\alpha=x,y}\sum_{\beta=x,y}
Q_{\alpha\beta}E_{\alpha}^{(inc)}E_{\beta}^{(inc)\,*}\, ,
\label{eq:eff}\\
Q_{\alpha\beta}=
(kR_1)^{-2}\pi^{-1}
\sum_{jm}
\left[
\beta_{jm;\,\alpha}^{(sca)}\beta_{jm;\,\beta}^{(sca)\,*}
+ n^{-2}
\tilde\beta_{jm;\,\alpha}^{(sca)}
\tilde\beta_{jm;\,\beta}^{(sca)\,*}
\right]\, ,
\label{eq:eff_comp}\\
\begin{pmatrix}
\beta_{jm;\,\alpha}^{(sca)}\\
n^{-1}\, 
\tilde{\beta}_{jm:\,\alpha}^{(sca)}
\end{pmatrix}
=\sum_{j'} \vc{T}_{jj';\,m}
\begin{pmatrix}
\alpha_{j'm;\,\alpha}^{(inc)}\\
n^{-1}\, 
\tilde{\alpha}_{j'm;\,\alpha}^{(inc)}
\end{pmatrix}\, .
  \label{eq:matr_uni}
\end{gather}
\end{widetext}
Note that for radial anisotropy
$Q_{\alpha\beta}=\delta_{\alpha\beta}\,Q$.  By contrast, in the case
of uniformly anisotropic droplet we have rather strong dependence of
the scattering efficiency on the polarisation of the incoming wave.
When the refractive indices are matched, $n=n_o$, it is expected that
the scatterer does not change the $y$ component of the incident wave,
which simply transforms into the ordinary wave inside the droplet
without being affected by the scattering process.  The algebraic
interpretation of this fact is that the amplitudes of the scattered
wave $\beta_{jm;\,y}^{(sca)}$ and $\tilde\beta_{jm;\,y}^{(sca)}$ are
equal to zero.  However, it is not straightforward to see that the
system~\eqref{eq:sys_fin} is consistent with this conclusion. We show
this in Appendix~\ref{sec:append}; the proof involves using
relations~(46.I) in which the electric and magnetic fields inside the
droplet are expressed in terms of appropriate normal modes.

\section{Numerical results}
\label{sec:num-res}
In this section we present numerical results for the scattering
efficiency defined by Eqs.~\eqref{eq:eff} and~\eqref{eq:eff_comp}. We
are primarily interested in anisotropy-induced scattering. In order to
concentrate on this test case, we consider the case when the
refractive indices $n$ and $n_o$ are equal.  We shall present a more
comprehensive analysis of all possible cases, including the results
for the angular distribution of the scattered waves, elsewhere.  We
begin with brief comments on numerical procedure and then proceed with
the description of the calculated dependencies.

It is rather straightforward to
perform computations for radially anisotropic droplets.
The expressions for the elements of \T-matrix are
known and given by Eqs.~\eqref{eq:t11_rad}-\eqref{eq:t22_rad}.
We can thus  evaluate the scattering efficiency by explicitly computing 
the sum in the expression~(22.I).

For uniformly anisotropic droplets
\T-matrix can only be computed numerically by solving
the system of equations~\eqref{eq:sys_fin}~\footnote{%
  These computations have been
  performed by using the NAG FORTRAN library at the University of Southampton.
  The library was employed to calculate special functions, evaluate integrals
  and inverse matrices. In order to have the relative error well below 0.1\%,
  the program was designed to take into account sufficiently large amount of
  contributions to the scattering efficiency that come from different angular
  momenta and azimuthal numbers. The detailed description of the program is
  beyond the scope of this paper. The code is available on request by e-mail
  from either of the authors.
}. Some highlights 
of the results are presented below.

The dependence of the scattering efficiency on the angle of incidence is shown
in Fig.~\ref{fig:uni_ang_u}. If the size parameter,
$kR_1$, is not very large, the scattering efficiency $Q_{xx}$ is a
monotonically increasing function of the angle of incidence,
$\theta_{inc}$, in the region from 0 to $\pi/2$. By symmetry
$Q_{xx}(\theta_{inc})=Q_{xx}(\pi/2-\theta_{inc})$, and  so
the scattering efficiency decreases in the range
from $\pi/2$ to $\pi$.

In Fig.~\ref{fig:uni_ang_kr} we show what happens for shorter
wavelength and thus higher values of $kR_1$.  Now, for for relatively
large values of the size parameter, the cross-section dependence on
the angle is incidence is no longer monotonic.  For example, at
$kR_1=20.0$, the angle at which the scattering efficiency $Q_{xx}$
reaches its maximum value is no longer at $\pi/2$.

Fig.~\ref{fig:eff_vs_kr} shows the scattering efficiencies
$Q_{xx}(\pi/2)$ (for uniform anisotropy) and $Q$ (for radial
anisotropy) versus the size parameter.  The scattering efficiency of
uniformly anisotropic droplet has a pronounced peak located at about
$kR_1\approx 10.0$ and exhibits strongly non-monotonic behaviour.  By
contrast, the corresponding dependence for the radially anisotropic
droplet is monotonically increasing. In this case the first maximum is
reached at $kR_1\approx 20.0$, outside the range of $kR_1$ shown in
Fig.~\ref{fig:eff_vs_kr}.

The scattering efficiencies as a function of the anisotropy
parameter, $0 \le u\le 1$, 
at different values of the size parameter 
are plotted in
Fig.~\ref{fig:eff_u_uni} and Fig.~\ref{fig:eff_u_rad} for
the cases of radial and uniform anisotropies respectively.
In both cases an increase in the size parameter leads to
the appearance of peaks in this range of $u$.
As compared to radially anisotropic scatterers, the uniformly anisotropic
droplets seem to be more sensitive to changes both in
the  size and in the anisotropy parameters .   
 
\section{Conclusion}
\label{sec:concl}
In this work we have described some of 
the numerical results calculated using the \T-matrix theory 
developed in paper~I.
In particular, we have studied the scattering efficiency
of radially and uniformly anisotropic droplets in which the ordinary
refractive index matches the refractive index of the material surrounding them.

The assumption in which the ordinary refractive index of the droplet
matches the isotropic dielectric constant in the surrounding medium is
not taken in order to simplify the numerical treatment. Rather in this
paper we wish to study the light scattering properties which can be
solely attributed to the \emph{anisotropic} part of the dielectric
tensor. Thus we have the anisotropy effects separated out to
concentrate on differences between isotropic and anisotropic optical
axis distributions.

Clearly, the difference in symmetry causes the most crucial
differences in the light scattering. For uniformly anisotropic droplets
the scattering efficiency depends on the angle of incidence and the
polarisation of incoming wave, whereas for radially anisotropic
scatterers it does not.  In other words, the scattering from radially
anisotropic droplet shares some features of isotropic scatterers.

Nevertheless, the angular dependencies for the scattered wave
intensity and the depolarisation factor shown at the end
of~I clearly indicate the pronounced differences.

The graphs plotted in Figs.~\ref{fig:eff_vs_kr}-\ref{fig:eff_u_rad}
indicate that uniformly anisotropic droplets are more sensitive to
changes in the wavelength and anisotropy parameters than are radially
anisotropic droplets.  Our results are also consistent with results of
previous studies~\cite{Zum:1988,Kis:2000:opt} that the internal
spatial distribution of the optical axis is a factor which strongly
affects light scattering from anisotropic scatterers.

The results of this work can be regarded as the first step towards
more comprehensive study of light scattering by anisotropic
scatterers.  We have demonstrated that the \T-matrix
approach developed in~I can be used in an efficient numerical
treatment of the scattering problem. It is thus reasonable to
expect that further progress can be made by applying this theory 
to more complex problems involving light scattering by optical 
anisotropic liquid crystalline systems and other related problems, 
as discussed further in the last section of~I.

\begin{acknowledgments}
We acknowledge support from INTAS under grant 99--0312. AK thanks the
Faculty of Mathematical Studies in the University of Southampton for
its hospitality for a number of visits during 2000 and 2001.
\end{acknowledgments}

\appendix
\section{}
\label{sec:append}
In this appendix we show mathematically, suing our formalism, the
physically obvious result that if the ordinary refractive index 
of a droplet matches that of the scattering medium, then there will be
no scattering of the polarisation component out of the plane of the 
incident wave and the uniform anisotropy in the droplet. In order to do this, 
we first  extend algebraic relations
that follow from Eqs.~(46.I). These equations give the expansion
of plane wave propagating in a uniformly anisotropic medium.
We can rewrite them for the plane wave inside the droplet:
\begin{widetext}
\begin{align}
&
\sum_{j'm'}
\left[\,
\alpha_{j'm'}^{(inc)}\,\vc{P}^{(m)}_{j'm'}(\rho_o,\uvc{r})
- n_o^{-1}\,
\tilde\alpha_{j'm'}^{(inc)}\,\vc{P}^{(e)}_{j'm'}(\rho_o,\uvc{r})
\,\right]=\notag\\
&
=
\exp(i\rho_e\,\uvc{k}_{inc}\cdot\uvc{r})
E_x(\uvc{k}_{inc})\bigl[
\vc{e}_x(\uvc{k}_{inc})+\frac{u}{1+u}\,\sin\theta_{inc}
\,\uvc{z}\bigr]+
\exp(i\rho_o\,\uvc{k}_{inc}\cdot\uvc{r})
E_y(\uvc{k}_{inc})\,\vc{e}_y(\uvc{k}_{inc})\, ,
\label{eq:A.1}\\
&
\sum_{j'm'}
\left[\,
\tilde\alpha_{j'm'}^{(inc)}\,\vc{Q}^{(m)}_{j'm'}(\rho_o,\uvc{r})
+n_o\,
\alpha_{j'm'}^{(inc)}\,\vc{Q}^{(e)}_{j'm'}(\rho_o,\uvc{r})
\,\right]=\notag\\
&
= n_o\left(\,
\exp(i\rho_e\uvc{k}_{inc}\cdot\uvc{r})\,
m_e^{-1}\,E_x(\uvc{k}_{inc})\vc{e}_y(\uvc{k}_{inc})-
\exp(i\rho_o\,\uvc{k}_{inc}\cdot\uvc{r})
E_y(\uvc{k}_{inc})\vc{e}_x(\uvc{k}_{inc})
\,\right)\, ,
\label{eq:A.2}
\end{align}
\end{widetext}
where $\rho_e=m_e\rho_o$ and
$m_e=\sqrt{(1+u)/(1+u\cos^2\theta_{inc})}$.
The coefficients $\alpha_{jm}^{(inc)}$
and $\tilde{\alpha}_{jm}^{(inc)}$ are defined by Eqs.~(9.I)
where the factor $\mu/n$ is changed to $1/n_o$.

We can now combine the relations
that come from definitions of the coefficient functions
(see Eq.~(45.I))
\begin{widetext}
\begin{align}
&
\sum_{j'\ge |m|}
\left[
p_{jj';\,m}^{\,(\alpha,m)}(\rho_o)\, 
\alpha_{j'm}^{(inc)}
- n_o^{-1}
p_{jj';\,m}^{\,(\alpha,e)}(\rho_o)\,
\tilde{\alpha}_{j'm}^{(inc)}
\right] =\notag\\
&
=
\Bigl\langle\,
\vc{Y}_{jm}^{(\alpha)\,*}(\uvc{r})\cdot
\sum_{j'm'}
\left[\,
\alpha_{j'm'}^{(inc)}\,\vc{P}^{(m)}_{j'm'}(\rho_o,\uvc{r})
- n_o^{-1}\,
\tilde\alpha_{j'm'}^{(inc)}\,\vc{P}^{(e)}_{j'm'}(\rho_o,\uvc{r})
\,\right] 
\,\Bigr\rangle_{\uvc{r}}\, ,
\label{eq:A.3}\\
&
\sum_{j'\ge |m|}
\left[
n_o\,q_{jj';\,m}^{\,(\alpha,e)}(\rho_o)\, 
\alpha_{j'm}^{(inc)}
+
q_{jj';\,m}^{\,(\alpha,m)}(\rho_o)\,
\tilde{\alpha}_{j'm}^{(inc)}
\right] =\notag\\
&
=
\Bigl\langle\,
\vc{Y}_{jm}^{(\alpha)\,*}(\uvc{r})\cdot
\sum_{j'm'}
\left[\,
\tilde\alpha_{j'm'}^{(inc)}\,\vc{Q}^{(m)}_{j'm'}(\rho_o,\uvc{r})
+n_o\,
\alpha_{j'm'}^{(inc)}\,\vc{Q}^{(e)}_{j'm'}(\rho_o,\uvc{r})
\,\right]
\,\Bigr\rangle_{\uvc{r}}\, ,\; \alpha\in\{m,e\}\, ,
\label{eq:A.4}
\end{align}
\end{widetext}
with the relations~\eqref{eq:A.1}-\eqref{eq:A.2} to evaluate   
the left hand side of the system~\eqref{eq:sys} 
provided that
$\{\alpha_{jm}^{(c)},\tilde{\alpha}_{jm}^{(c)}\}=
\{\alpha_{jm}^{(inc)},\tilde{\alpha}_{jm}^{(inc)}\}$.

To this end, we can use Eq.~\eqref{eq:r_uni} to write down
the sum on the left hand side of Eq.~\eqref{eq:sys} in the following
form:
\begin{widetext}
\begin{equation}
\sum_{j'\ge |m|}\vc{R}^{jj';\,m}(r)
\begin{pmatrix}
\alpha_{j'm}^{(inc)}\\
\tilde{\alpha}_{j'm}^{(inc)}
\end{pmatrix}
 =\sum_{j'\ge |m|}
\begin{pmatrix}
p_{jj';\,m}^{\,(m,m)}(\rho_o)\, 
\alpha_{j'm}^{(inc)}
- n_o^{-1}
p_{jj';\,m}^{\,(m,e)}(\rho_o)\,
\tilde{\alpha}_{j'm}^{(inc)}\\
n_o\,q_{jj';\,m}^{\,(e,e)}(\rho_o)\, 
\alpha_{j'm}^{(inc)}
+
q_{jj';\,m}^{\,(e,m)}(\rho_o)\,
\tilde{\alpha}_{j'm}^{(inc)}\\
n_o\,q_{jj';\,m}^{\,(m,e)}(\rho_o)\, 
\alpha_{j'm}^{(inc)}
+
q_{jj';\,m}^{\,(m,m)}(\rho_o)\,
\tilde{\alpha}_{j'm}^{(inc)}\\
p_{jj';\,m}^{\,(e,m)}(\rho_o)\, 
\alpha_{j'm}^{(inc)}
- n_o^{-1}
p_{jj';\,m}^{\,(e,e)}(\rho_o)\,
\tilde{\alpha}_{j'm}^{(inc)}
\end{pmatrix}\, .
\label{eq:A.5}
\end{equation}
\end{widetext}
It is seen that the elements of the column on the right hand side
of this equation are the sums from the left hand sides
of Eqs.~\eqref{eq:A.3} and~\eqref{eq:A.4}.
On the other hand, from Eqs.~\eqref{eq:A.1}-\eqref{eq:A.2}, 
the square bracketed sums on the right hand sides of
Eqs.~\eqref{eq:A.3}-\eqref{eq:A.4} are the plane waves.
So, the elements of the column~\eqref{eq:A.5} can be evaluated as
scalar products of the vector spherical functions and the vector plane
waves by using Eqs.~(B6.I) and~(B8.I) of appendix~B in~I.

We can now apply this procedure  to calculate the elements
of the column~\eqref{eq:A.5}
for the ordinary wave with 
$\{\alpha_{jm}^{(inc)},\tilde{\alpha}_{jm}^{(inc)}\}=
\{\alpha_{jm;\,y}^{(inc)},\tilde{\alpha}_{jm;\,y}^{(inc)}\}$.
From Eqs.~\eqref{eq:alp}-\eqref{eq:talp} we have
$E_x(\uvc{k}_{inc})=0$
and $E_y(\uvc{k}_{inc})=1$ in this case. The final result is
\begin{widetext}
\begin{equation}
\sum_{j'\ge |m|}\vc{R}^{jj';\,m}(r)
\begin{pmatrix}
\alpha_{j'm;\,y}^{(inc)}\\
\tilde{\alpha}_{j'm;\,y}^{(inc)}
\end{pmatrix}
 =
\alpha_{jm;\,y}^{(inc)}
\begin{pmatrix}
j_j(\rho_o)\\
n_o\,[j_j(\rho_o)]'\\
0\\
0
\end{pmatrix}
+n_o^{-1}\,
\tilde{\alpha}_{jm;\,y}^{(inc)}
\begin{pmatrix}
0\\
0\\
n_o\,j_j(\rho_o)\\
-[j_j(\rho_o)]'
\end{pmatrix}\, .
\label{eq:A.6}
\end{equation}
\end{widetext}

When $n_o=n$ (and $\rho=\rho_o$),
after multiplying~\eqref{eq:A.6} by the matrices~\eqref{eq:aux_matr},
we have 
\begin{widetext}
\begin{equation}
\sum_{j'\ge |m|}\vc{B}_{jj';\,m}
\begin{pmatrix}
\alpha_{j'm;\,y}^{(inc)}\\
\tilde{\alpha}_{j'm;\,y}^{(inc)}
\end{pmatrix}=
\begin{pmatrix}
\alpha_{jm;\,y}^{(inc)}\\
n^{-1}\,\tilde{\alpha}_{jm;\,y}^{(inc)}
\end{pmatrix}\, ,\quad
\sum_{j'\ge |m|}\vc{A}_{jj';\,m}
\begin{pmatrix}
\alpha_{j'm;\,y}^{(inc)}\\
\tilde{\alpha}_{j'm;\,y}^{(inc)}
\end{pmatrix}=
\begin{pmatrix}
0\\
0
\end{pmatrix}\, .
\label{eq:A.7}
\end{equation}
\end{widetext}
From these equations 
we immediately conclude that, when
$n=n_o$ and $\{\alpha_{jm}^{(inc)},\tilde{\alpha}_{jm}^{(inc)}\}=
\{\alpha_{jm;\,y}^{(inc)},\tilde{\alpha}_{jm;\,y}^{(inc)}\}$,
the solution of the system~\eqref{eq:sys_fin} 
is given by
\begin{widetext}
\begin{equation}
  \label{eq:A.8}
 \alpha_{jm}^{(c)}=
\alpha_{jm;\,y}^{(inc)},\quad
\tilde{\alpha}_{jm}^{(c)}=\tilde{\alpha}_{jm;\,y}^{(inc)},\quad
\beta_{jm}^{(sca)}\equiv\beta_{jm;\,y}^{(sca)}=0,\quad
\tilde\beta_{jm}^{(sca)}\equiv\tilde\beta_{jm;\,y}^{(sca)}=0\,. 
\end{equation}
\end{widetext}
So, the amplitudes of scattered wave 
$\beta_{jm;\,y}^{(sca)}$ and $\tilde\beta_{jm;\,y}^{(sca)}$ 
vanish at $n=n_o$. 


\end{document}